\documentclass[useAMS,usenatbib,a4paper, fleqn]{mnras}
\usepackage{ae,aecompl}
\title[Young massive cluster formation]
{The formation of young massive clusters by colliding flows
}
\author[Dobbs]
{C. L. Dobbs\thanks{E-mail:
C.L.Dobbs@exeter.ac.uk}$^{1}$, K. Y. Liow$^{1}$. S. Rieder$^{1}$ \\
$^1$ School of Physics and Astronomy, University of Exeter, Stocker Road, Exeter, EX4 4QL, UK \\
}

\usepackage{amssymb}
\usepackage{amsmath}
\usepackage[pdftex]{graphicx}
\usepackage{epsfig}
\usepackage{multirow} 
\def\aap{{ A\&A}}

\def\apjl{{ApJL}}

\def\apjs{{ApJS}}

\begin{document}
\label{firstpage}
\date{\today}

\pagerange{\pageref{firstpage}--\pageref{lastpage}} \pubyear{2019}

\maketitle

\begin{abstract}
Young massive clusters (YMCs) are the most intense regions of star formation in galaxies. Formulating a model for YMC formation whilst at the same time meeting the constraints from observations is highly challenging however. We show that forming YMCs requires clouds with densities $\gtrsim$ 100 cm$^{-3}$ to collide with high velocities ($\gtrsim$ 20 km s$^{-1}$).
We present the first simulations which, starting from moderate cloud densities of $\sim100$ cm$^{-3}$, are able to convert a large amount of mass into stars over a time period of around 1 Myr,  to produce dense massive clusters similar to those observed. Such conditions are commonplace in more extreme environments, where YMCs are common, but atypical for our Galaxy, where YMCs are rare.  
\end{abstract}
\begin{keywords}
galaxies: ISM, ISM: clouds, stars: formation, galaxies: star clusters: general
\end{keywords}

\section{Introduction}

Young massive clusters (YMCs) are the densest, most massive star clusters that are still forming in the present day \citep{PZ2010}. They are relatively rare in our Galaxy, but common in some other environments such as the Antennae which is undergoing a galaxy merger. YMCs are characterised by higher densities compared to open clusters, masses of $\gtrsim 10^4$ M$_{\odot}$, and also exhibit very short age spreads of the order of 1 Myr \citep{Longmore2014}. These properties represent a significant challenge for their formation - in short, one requires a very large mass of gas to be gathered in a small region of space on a very small timescale. 

The typical picture of star cluster formation in astrophysics is of a turbulent giant molecular cloud which collapses under gravity. For molecular cloud densities of around 100 cm$^{-3}$, the free fall time, i.e. the timescale for the collapse of the cloud, is around a few Myr. Over this timescale, one or multiple stellar groups or clusters can form within the cloud with slightly different stellar ages and age spreads. The molecular cloud can form via potentially any one of a number of processes which convert non-star forming cold atomic gas to molecular gas (see \cite{Dobbs2014}).

There are a few potential issues with this process for the formation of YMCs from isolated clouds. Firstly, simply the timescales may be too long compared to observations for collapse on a free fall time alone. Secondly, particularly for filamentary clouds, star formation may not be concentrated into a single massive cluster. Thirdly, molecular clouds are not readily observed which are not undergoing star formation (outside the Galactic Centre), so it is unclear how an isolated molecular cloud would come into existence without undergoing strong star formation, and then suddenly collapse. Finally if clouds simply collapsed to form YMCs under gravity, then we would expect YMCs to occur everywhere in the Galaxy, which is not the case. Such arguments favour a `conveyor belt' model of formation, whereby gas is continually accreted onto a forming cluster \citep{Longmore2014, Krumholz2019} in an atypical location rather than formation from an isolated cloud.
 
\section{Colliding clouds as a mechanism for forming YMCs}
One way of gathering gas together in a shorter timescale is by colliding flows of gas with large velocities. This mechanism has the advantage that YMC formation is likely to be promoted in merging galaxies (see also \cite{Jog1992}), where indeed YMCs are common, and be less favourable in more quiescent environments like the Galaxy. 

\begin{figure}
\centerline{
\includegraphics[scale=0.4, bb=0 180 570 640]{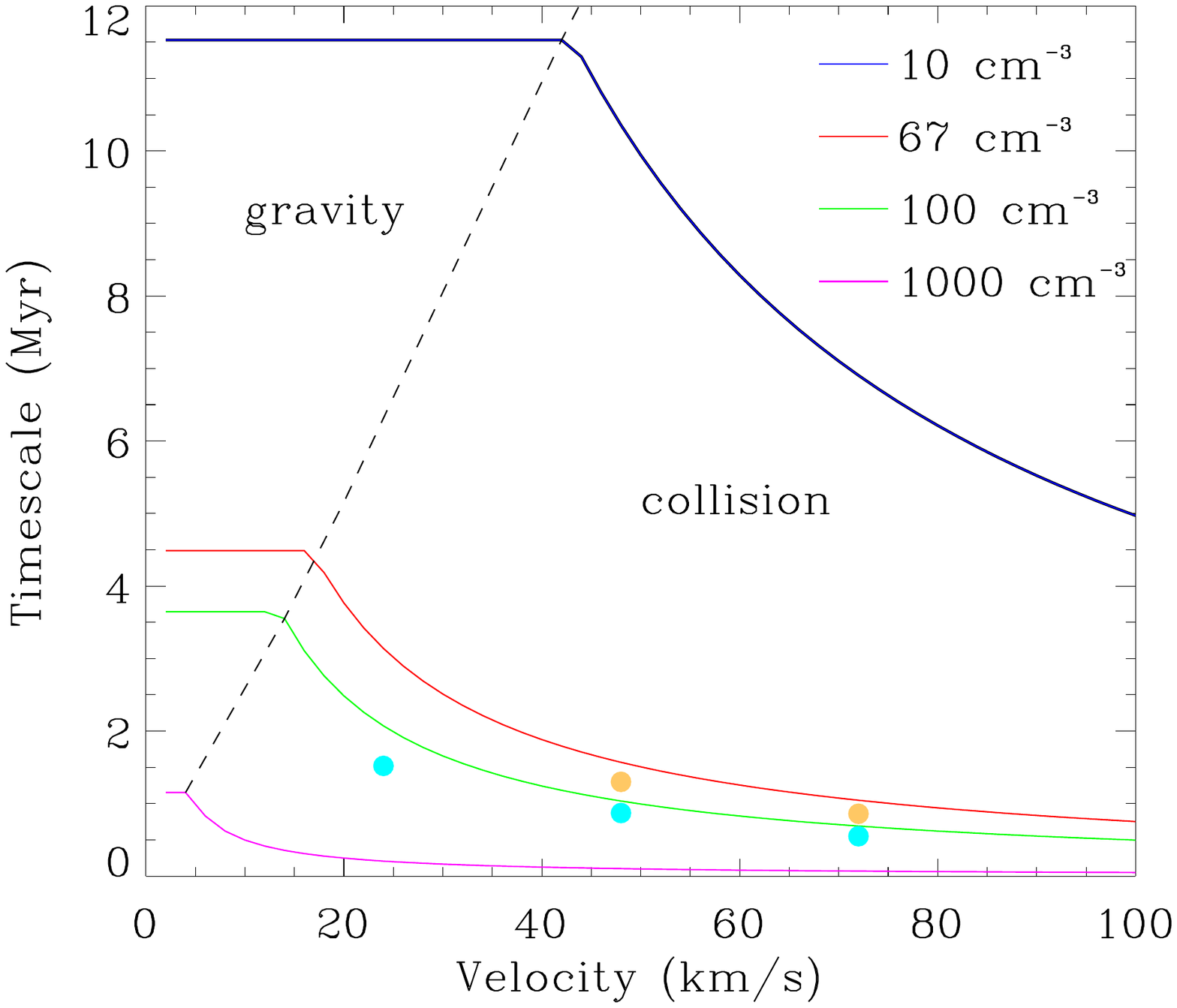}}
\caption{The timescale for forming a $2\times10^4$ M$_{\odot}$ cluster is shown where the timescale is the minimum of the free fall time (dependent only on density) and the timescale from two colliding flows where the relative velocity is plotted along the $x$ axis. The lines show theoretical values for different densities, and the orange (low density, 67 cm$^{-3}$) and cyan (standard density, 100 cm$^{-3}$) points are from the simulations. The figure shows two regimes, one where the collision has no effect and only gravity is significant, and a regime (right hand side) where the cluster will form faster due to the collision than gravity alone. A timescale of $\lesssim 2$ Myr for YMC formation implies that except at very high densities, collisions are likely to be required.}
\label{fig:timescale}
\end{figure}

We compare cluster formation by colliding flows with that from isolated clouds, and also predict the regimes where colliding flows are important. In Figure~\ref{fig:timescale} we show the timescale to form a cluster of mass $2 \times 10^4$ M$_{\odot}$, assuming this timescale is the minimum of the time either to form by a cloud collision, or simply through gravitational collapse alone (\citet{Krumholz2019} perform similar analysis for clusters, but do not include collisions). 
The latter is simply the free fall time, $t_{ff}=\sqrt{\frac{3 \pi}{32 G \rho}}$, which is only dependent on density. The timescale for the collision is calculated as follows.
We estimate how much gas can be accumulated into a GMC via a cloud collision similar to \citet{Pringle2001}.  To a simple approximation, we can estimate the mass of a cluster forming from a collision from the mass of gas which enters the shock (this is accurate assuming the shock is supersonic, Liow \& Dobbs. in prep.). This is just $\sim \rho_0 A v_o t_{col}$ where $\rho_0$ is the initial gas density, $A$ is the cross sectional area, $v_0$ the initial velocity flow and $t$ the time. We can then convert this into an approximate cluster mass by adopting a constant star formation efficiency $\epsilon$, so $M_{cluster}\sim\epsilon \rho_0 A v_o t_{col}$.  We then rearrange to find the timescale $t_{col}$. For Figure~\ref{fig:timescale}, we choose dimensions similar to those in our simulations, and take an efficiency of 20 \%. This efficiency is consistent with observed estimates for dense gas (10-30\%; \citealt{Lada2003}).

As Figure~\ref{fig:timescale} indicates, the gas needs to be both suitably dense ($>$ 100 cm$^{-3}$), and the clouds undergoing a high velocity ($>$ 20 km s$^{-1}$), to form a YMC, and the higher the gas densities, and collision velocities, the more likely the gas is to form a YMC. With low densities, the timescales are too long (either for self-gravity to form stars, or enough gas to be assembled) for YMC formation. With lower velocities, the gas cannot be converted to star-forming regions on a short enough timescale, and / or the gas starts to form stars but on a longer timescale.   

\section{Simulations of YMC formation}
We now present results from simulations investigating the possible formation of YMCs under different conditions. Here we are concerned predominantly with which conditions YMC formation is possible under, and do not model the full physics such as magnetic fields and stellar feedback. Moreover we assume that the clusters form so quickly that stellar feedback has not had time to significantly influence the gas (the simulations by \cite{Howard2018} show that feedback has minimal impact for well over 1 Myr). We list the different calculations performed in Table~\ref{tab:models}.

\begin{table*}
\begin{tabular}{c|c|c|c|c|c}
 \hline 
Model & Density & Velocity &  Time to form   & Mass of cluster & R$_{eff}$ \\
 & (cm$^{-3}$) &  (km s$^{-1}$) &  $2\times10^4$ M$_{\odot}$  stars & found ($10^3$ M$_{\odot}$) & (pc) \\
Collision (fiducial) & 100 & 48 & 0.88  & 8.9 & 0.9 \\
Collision (low velocity) & 100 &  24 & 1.5 & 12 & 2.2\\
Collision (high velocity) & 100 & 72 & 0.55 & 2.1 & 2.5 \\
No Turbulence & 100 & 48 & 0.35 & 10.0 & 4.6 \\
Isolated & 100 & 0 & 1.8 & 2.1 & 0.5 \\
Isolated shear & 100 & 0 & 2.0 & 1.3 & 1.6 \\
Turbulent box & 100 & 0 & 3.3 & 6.4 & 0.1  \\
Low density collision & 67 & 48 & 1.3 & 4.6 & 2.4 \\
Low density collision & 67 & 72 & 0.85  &  2.5 & 2.7  \\
Low density isolated & 67 & 0 & 2.7 & 4.1 & 0.8 \\
\hline
\end{tabular}
\caption{List of simulations performed. The velocity represents the velocity of the collisions, and the time is to form $2\times10^4$ M$_{\odot}$ of stars. The mass of cluster represents the largest mass cluster picked out with DBSCAN, and the $R_{eff}$ is the half mass radius of this cluster. The collision velocities represent the relative velocities between the two clouds. Turbulent driving has similar densities and turbulence to the fiducial simulation. Times represent the amount of time which has elapsed since star formation started. Typically star formation does not begin until $\sim1$ Myr, or longer in the isolated and turbulent box cases.}
\label{tab:models}
\end{table*}

We performed these simulations using phantom \citep{Price2018} which is a publicly available Smoothed Particle Hydrodynamics code. All calculations use 5 million particles. For most cases we set up ellipsoidal clouds with a length in one dimension of 80 pc along the $x$ axis, and 20 pc in the other two dimensions (we choose elongated clouds partly since molecular clouds are typically supposed to be filamentary). The clouds collide along one of the two shorter axes. If the clouds are less elongated, then the material along the axis of the collision has finished entering the shock before a cluster has time to develop. The clouds have masses of $1.5 \times 10^5$ M$_{\odot}$ in the fiducial case, and $10^5$ M$_{\odot}$ in the low density case. The densities in these two cases are then $\sim$ 100 cm$^{-3}$ and $\sim$ 67 cm$^{-3}$ respectively. A turbulent velocity field is added as described in \citet{Bate2003}. The turbulent velocity dispersion is $\sim 6$ km s$^{-1} $ in all calculations with turbulence. The cloud the kinetic and potential energies are initially similar, with the kinetic energy around 1.5 times that of the potential energy.  We also model an isolated cloud subject to a galactic potential (`Isolated shear'). All the simulations except the `No Turbulence' model include turbulence.  For the `No turbulence' and `Turbulent box' models, the particles are initially distributed with a close packed structure, within boxes of dimensions $32 \times 16 \times 16$ pc and (30 pc)$^3$ respectively. For `Turbulent box' model, turbulence is instead driven throughout the simulation as described in \citet{Price2010} and produces a similar velocity dispersion to the non-driven cases (at early times in the latter).  

The clouds are assumed to be molecular, and isothermal (20 K). Our densities represent low density molecular clouds, or high density atomic clouds. For the latter, the initial temperature of the gas would be higher (50--100 K), however we would still expect similar results (and indeed if we increase the temperature to 100 K our simulations produce similar results). As our analysis of the cloud collisions does not include the sound speed, and instead assumes that the collisions are strongly supersonic, this will still be true if the gas is cold HI, and consequently a strong shock will still develop. We note that an isothermal equation of state may also suppress instabilities present in adiabatic cases (e.g. \citealt{Nakamura2006,Goldsmith2020}).

We insert sink particles according to criteria in \citet{Bate1995}, adopting a critical density of $10^{-18}$ g cm$^{-3}$ and an accretion radius of 0.01 pc. With our resolution, we cannot model individual stars, rather each sink particle typically represents a small group of stars. Mergers between sink particles do not occur. Artificial viscosity is included with a switch for the $\alpha$ parameter \citep{Morris1997}. We choose $\beta=4$ in the colliding cloud cases, as recommended for strong shock \citep{Price2010}, and $\beta=2$ in the non-colliding cases. Varying $\beta$ has little effect on the non-colliding models, but using $\beta=4$ produces much less noisy shocks in the colliding cloud models.

\begin{figure}
\centerline{
\includegraphics[scale=0.32, bb=0 170 500 620]{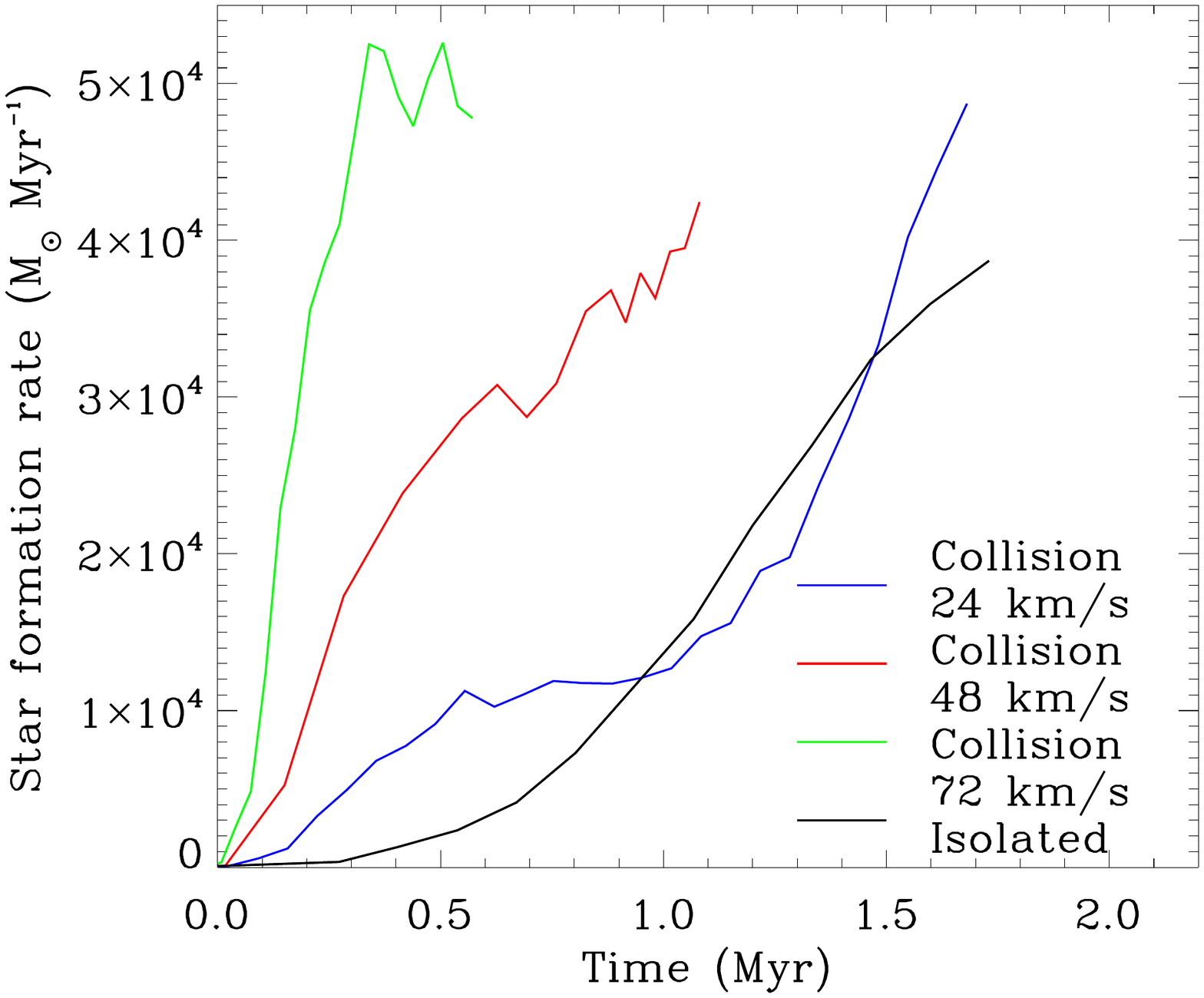}}
\caption{The star formation rates are shown versus time for the 100 cm$^{-3}$ density models for the colliding and isolated clouds. The times indicate the time since star formation commences.}
\label{fig:sfrate}
\end{figure}

For each simulation, we show in Table~\ref{tab:models} the time taken for masses of $2\times 10^4$ M$_{\odot}$ of stars to form. As indicated in Table~\ref{tab:models}, it is possible to form masses comparable to YMCs on timescales of 0.5-1.5 Myr with colliding flows. As expected, increasing the velocity of the collision and density of the gas increases the star formation rate (see also Figure~\ref{fig:sfrate}). The model which is most efficient at forming stars on a short timescale is the one without turbulence, which is closest to the theoretical picture in the previous section. However the stars are aligned in a 2D distribution rather than a sphere (the stars relax into a spherical cluster over $\sim 2$ Myr). 
 
 We also applied the clustering algorithm DBSCAN to the sink particles formed in the simulations. In Table~\ref{tab:models}, we list the masses and radii of the most massive clusters picked out using DBSCAN, adopting a maximum separation of 0.5 pc. This indicates the sizes and masses of sink particles clustered together rather than simply the total mass. As indicated in Table~1, a major limitation of most of the models is that the masses of the detected clusters are relatively low. The exceptions are the no turbulence, lower and fiducial velocity collision cases, where clusters of mass $10^4$ M$_{\odot}$ are formed. The downside of the collisional models (particularly the no turbulent case), is that the physical size of the cluster is initially dominated by the shape of the shock, and thus can be larger compared to the other models (see Liow \& Dobbs, in prep.). Moderate mass clusters are formed in the turbulent box model, but these also take the longest time to form.

In Figure~\ref{fig:sfrate} we plot the star formation rates for the standard density simulations. It is clear that the for the cases of the colliding clouds, the star formation rates increase more quickly than the corresponding isolated models, and reach higher values. As would also be expected, the star formation rates reach higher values for the higher velocity collisions.
\begin{figure}
\centerline{
\includegraphics[scale=0.35, bb=50 0 680 570]{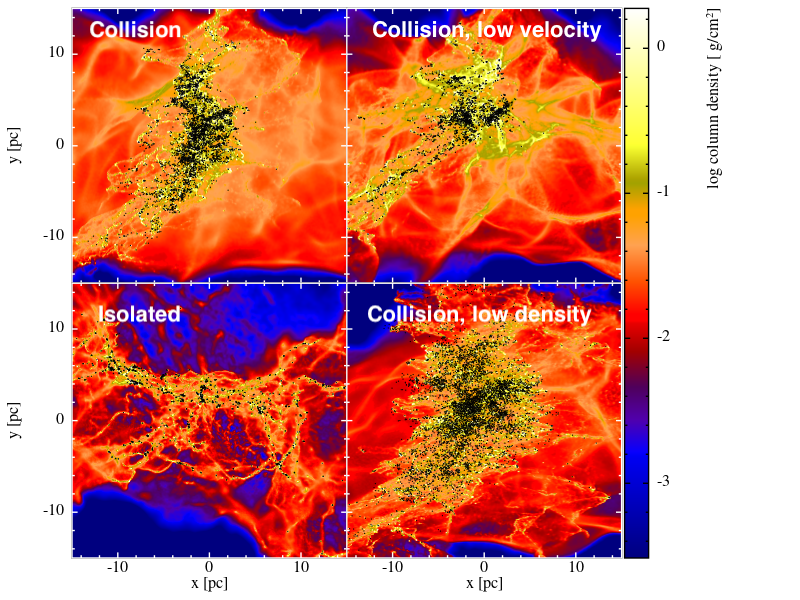}}
\caption{The column density maps are shown for the fiducial colliding, low velocity colliding, low density colliding and non-colliding simulations. The colour map shows the gas density, and the black dots represent sink particles. The panels are shown after a mass of $2\times 10^4$ M$_{\odot}$ of stars has formed.}
\label{fig:xy}
\end{figure}

\begin{figure}
\centerline{
\includegraphics[scale=0.35, bb=50 0 680 600]{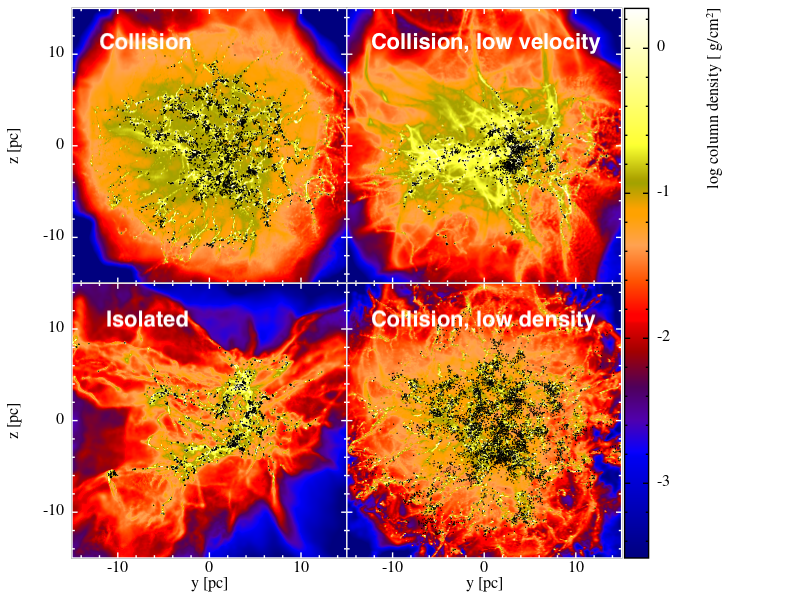}}
\caption{The column density maps are shown for the fiducial colliding, low velocity colliding, low density colliding and non-colliding simulations, here showing the $y$\textendash $z$ plane. The colour map shows the gas density, and the black dots represent sink particles. The panels are shown after a mass of $2\times 10^4$ M$_{\odot}$ of stars has formed.}
\label{fig:yz}
\end{figure}

\begin{figure}
\centerline{
\includegraphics[scale=0.45,  bb=0 80 460 500]{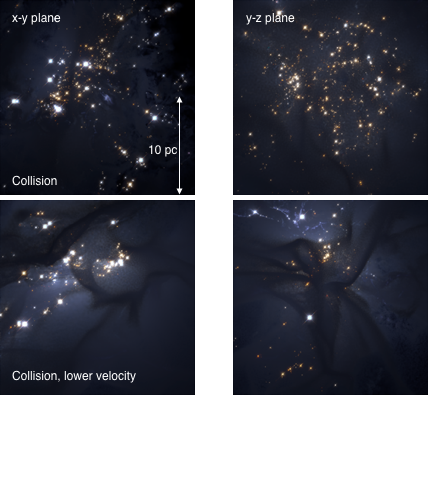}}
\caption{HST style images are shown for the fiducial collision, low velocity collision, low density collision, and isolated case using the fresco package (Rieder \& Pelupessy, available on GitHub). Fresco includes extinction from dust, calculated from the gas in the simulations assuming a constant dust to gas ratio.}
\label{fig:hst}
\end{figure}

We show the column density of four of the simulations in Figures~\ref{fig:xy} ($x$\textendash $y$) and \ref{fig:yz} ($z$\textendash $y$) plane. The structure of the fiducial collision is similar to \citet{Balfour2015}, showing fragmentation in the plane perpendicular to the shock. The lower velocity, and lower density collisions  show more concentrated clusters. The masses and radii of the clusters formed in the fiducial, and lower velocity colliding flow simulations are comparable to NGC 3603 in our Galaxy and lower mass YMCs of external galaxies though the latter tend to be much older (see Figure~2 and Table~3 of \cite{PZ2010}). Figure~\ref{fig:hst} shows synthetic HST style images to show how these would appear as observed clusters. The fiducial collision case shows a less compact cluster, although we find that over time, a clearer central more spherical cluster emerges. Here the initial cluster structure is shaped somewhat by the structure of the shock. For the lower velocity model, gravity has time to start acting by the time the equivalent amount of material has collided, and as such a more compact cluster has chance to form. For the lower velocity case, the gas builds up over a longer time, and the morphology of the stars has evolved further away from the shape of the shock compared to the standard density model. The isolated case instead shows separate distinct low mass (e.g. from Table~\ref{tab:models}) clusters.

\begin{figure}
\centerline{
\includegraphics[scale=0.35, bb=50 0 680 550]{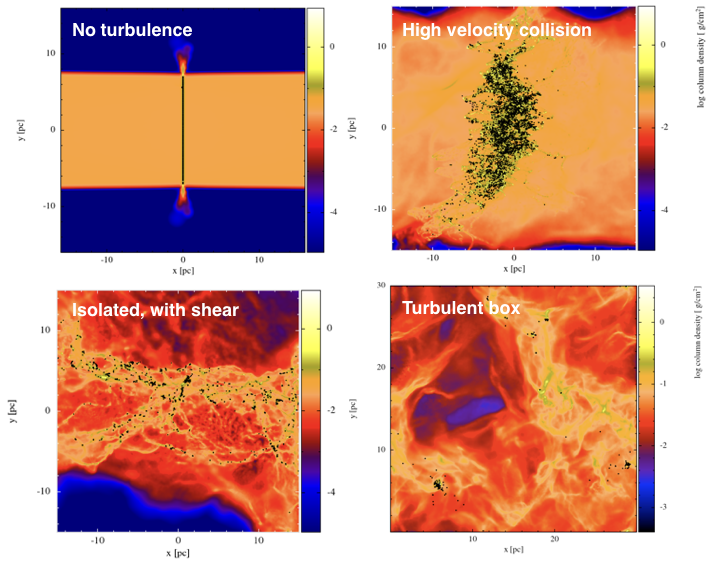}}
\caption{The column density maps are shown for the no turbulence, high velocity collision, isolated cloud subject to shear and turbulent box simulation. The colour map shows the gas density, and the black dots represent sink particles. The panels are shown after a mass of $2\times 10^4$ M$_{\odot}$ of stars has formed. For the isolated cloud subject to shear some of the star formation lies outside the region shown.}
\label{fig:other}
\end{figure}

In Figure~\ref{fig:other} we show column density images from further simulations. Without turbulence, the star formation occurs in a sheet morphology within a very short timescale. The level of turbulence can be considered a factor in the efficiency of star formation used in the earlier analysis. For no turbulence, the efficiency is much closer to 1, and the timescale correspondingly smaller. For the higher velocity collision, the morphology is very similar to the fiducial collision model, but simply occurs at an earlier timeframe. The isolated cloud with shear appears similar to the isolated case without shear, the clusters are simply more dispersed. Again there appear to be multiple smaller clusters in this example. Finally in the turbulent box model there are multiple clusters, which form at the sites of convergent flows in the turbulence.

The simulations support the expectation that high collision velocities are needed to have a significant impact on the star formation rate, and produce short formation timescales.  The velocities are required to be high compared to the sound speed, and timescales short compared to the free-fall time (Figure~1) in our analytic arguments and models. Additionally the velocities are high compared to the turbulence in the models otherwise the spatial distribution of stars tends to follow the turbulent structure rather than forming a dense cluster (see also Liow \& Dobbs, in prep.). The cloud collision models also clearly focus the dense gas into a confined region, which is not the case with the isolated clouds (or the turbulently driven box). This may be true to some extent even with lower collision velocities than those tested here. The outcome of the isolated models will depend to some extent on the geometry and turbulent velocity of the cloud, but of various realisations, we typically find that numerous smaller clusters are formed rather than one single massive cluster. This is a different outcome to \citet{Howard2018}, who conclude that a single cluster forms (as well as modelling elongated clouds, we also do not include mergers of sink particles which would decrease the resolution of the stellar component). 
In the likely more realistic cases where there is driven turbulence or galactic shear (or potentially magnetic fields), these increase the timescales. These processes would be less important or absent in the colliding cases (since the collision velocities are higher than turbulence, whilst collisions will occur at locations of converging rather than diverging flows).

\section{Discussion}
We have shown, both using theoretical arguments and numerical simulations, that it is possible to form massive clusters from colliding flows or clouds in timescales of $\lesssim1.5$ Myr. The conditions for YMCs to form are that the gas needs to be at least moderately dense ($>100$ cm$^{-3}$) and undergoing high ($>20$ km s$^{-1}$) collision velocities. Otherwise, the timescales to accumulate large masses of gas and turn the gas into stars are too long, and multiple smaller clusters form. The conditions from which the clusters form are atypical for the Milky Way but not implausible. Since YMCs are rare in the Milky Way, we would not expect the conditions from which they arise to occur frequently. We estimate the rate of collisions of GMCs to be 1 in every 8--10 Myr in \citet{Dobbs2015}, comparable to the lifetimes of the clouds, although the rate of collisions of the most massive clouds is significantly less than this. Collision velocities $>10$ km s$^{-1}$  are not uncommon, whilst the highest collision velocities are $\sim 20$ km s$^{-1}$ \citep{Furukawa2009,Motte2014,Fukui2015,Fukui2018}. Higher collision velocities are further likely with a stronger spiral potential (Rieder et al., in prep.), or at specific locations such as the end of the bar \citep{Motte2014}. Furthermore, higher surface densities such as in the more inner parts of the Galaxy, will lead to more collisions of high mass clouds. On the other hand, in the Antennae, where massive clusters are common, a velocity difference of 125 km s$^{-1}$ has been observed for one possible proto-globular cluster \citep{Finn2019}. Even though they modelled isolated clouds, \citet{Fujii2016} also concluded that cloud--cloud collisions are important to produce the high velocity dispersions used in their simulations of massive clouds.

Our results also show that there is a surprisingly small difference in terms of star formation rates between the isolated cases, and the colliding clouds, unless extreme velocities are used. Again this is in agreement with YMCs as more extreme occurrences. However the collision is also relevant in focusing gas together in the same region of space (which could be a spiral arm or a cloud colliding with the Galaxy, see also \cite{Alig2018}, or galaxy-galaxy collision), whereas in the isolated clouds there is no single central cluster. This becomes more apparent as shear is included, which even starting from a more spherical cloud, will still elongate the cloud. Likewise continuously driven turbulence may further reduce star formation (in the colliding clouds case the cluster forms before turbulence significantly decays which is not the case for the isolated clouds).  Either (and potentially also magnetic fields) may help explain why clouds in quiescent environments such as the Milky Way are not  generally collapsing to form massive clusters, or at least help to delay star formation until stellar feedback becomes effective.

We will present a resolution study of cluster formation in upcoming work (Liow \& Dobbs, in prep.) but note that we do not find significant differences in star formation rates, or the trends seen here with resolution above 1 million particles. In carrying out this work, we ran a few realisations, which suggest that all the timescales to form masses are likely subject to uncertainties on the order of 20 \% due to the turbulent field, and the exact nature of the collision. We leave the inclusion of magnetic fields and stellar feedback to future work.

\section{Acknowledgments}
CLD acknowledges funding from the European Research Council for the Horizon 2020 ERC consolidator grant project ICYBOB, grant number 818940. SR acknowledges funding from the STFC Consolidated Grant ST/R000395/1. Calculations for this paper were performed on the ISCA High Performance Computing Service at the University of Exeter, and the DiRAC DIAL system, operated by the University of Leicester IT Services, which forms part of the STFC DiRAC HPC Facility (www.dirac.ac.uk ). This equipment is funded by BIS National E-Infrastructure capital grant ST/K000373/1 and  STFC DiRAC Operations grant ST/K0003259/1. DiRAC is part of the National E-Infrastructure.  Figures in this paper were produced using \textsc{splash} \citep{splash2007}.

\bibliographystyle{mn2e}
\bibliography{Dobbs}
\bsp
\label{lastpage}
\end{document}